\documentclass[runningheads,a4paper]{llncs}

\usepackage[utf8]{inputenc}
\usepackage{amssymb}
\usepackage{graphicx}
\usepackage{multirow}
\usepackage{multicol}
\usepackage{subcaption}
\usepackage{fancyvrb}

\usepackage{url}
\newcommand{\keywords}[1]{\par\addvspace\baselineskip
\noindent\keywordname\enspace\ignorespaces#1}

\begin{document}


\mainmatter

\title{Transportation in Social Media: an automatic classifier for travel-related tweets}

\author{João Pereira$^{1,2}$, Arian Pasquali$^{3}$, Pedro Saleiro$^{1,2}$, Rosaldo Rossetti$^{1,2}$}
\institute{$^1$FEUP, $^2$LIACC, $^3$INESC TEC, Universidade do Porto, Portugal
  {\{joao.filipe.pereira,pssc,rossetti\}@fe.up.pt, arrp@inesctec.pt }
}
\maketitle

\begin{abstract}
In the last years researchers in the field of intelligent transportation systems have made several efforts to extract valuable information from social media streams. However, collecting domain-specific data from any social media is a challenging task demanding appropriate and robust classification methods. In this work we focus on exploring geo-located tweets in order to create a travel-related tweet classifier using a combination of bag-of-words and word embeddings. The resulting classification makes possible the identification of interesting spatio-temporal relations in São Paulo and Rio de Janeiro.  

\end{abstract}

\section{Introduction}

Social media data is still in the process of maturation regarding its use in the transportation and mobility fields. Users tend to publicly share events in which they participate, as well as the ones related to the operation of the transportation network, such as accidents and other disruptions. The exploration of social media data brings particular advantages, under virtually no cost, such as real-time data and content authenticity due to its human generated nature. In the domain of transportation, social media data analysis can produce valuable insights to support traffic management and control, human mobility, shared services, and policy making studies. 

The main goal of this work lies upon the development of an automatic system capable of discriminating travel-related tweets from a stream of geo-located tweets.
However, extracting accurate knowledge from social media content is still challenging. Social media texts are usually short, informal, with a lot of abbreviations, jargon, slang and idioms, being such characteristics the biggest challenges to surpass regarding text analysis.

We take advantage of recent developments on neural language models \cite{mikolov2013distributed}, and train continuous word representations, the so called word embeddings, from two streams of geo-located tweets in São Paulo and Rio de Janeiro, comprising more than 8 millions tweets. These word embeddings representations are able to capture syntactic and semantic relations between words in tweets that are useful in the context of transportation (e.g. ``ônibus'' and ``busão''). 

Consequently, we combine bag-of-embeddings and standard bag-of-words to learn a binary classifier of travel/non-travel tweets classification. We train and evaluate our model using a new hand-coded gold standard specifically created for this work. We applied the resulting classifier to both streams of geo-located tweets and obtained interesting spatio-temporal relations on both cities, demonstrating the usefulness of travel-related filtering on streams of geo-located tweets.

The remainder of this paper is organised as follows. In Section \ref{related_work}, we present some of the related work already done in the area. The methodology behind the collection of geo-located tweets, as well as, its exploratory analysis are detailed in Section \ref{data}. In Section \ref{classifier}, we describe our approach, the processing steps applied to our data and the features selection. The experimental set up is presented in Section \ref{exp_setup}, while the results of our experiences and their discussion are presented in Section \ref{results}. We draw conclusions and highlight main contributions  in Section \ref{conclusions_future_work}, alongside some suggestions for further developments in this project.

\section{Related Work}\label{related_work}

Classification models applied to Twitter data were already proposed and tested in different areas of study, ranging from sentiment analysis for political data science \cite{saleiro2015popmine,saleiro2016sentiment} to predict stock market fluctuations \cite{saleiro2017feup}, or to predict content popularity on Twitter \cite{saleiro2016learning}. Recently, the research community has realised the potential of using social media data to study the various dimensions of transportation systems, including mobility, travel purpose and different modes of transport and their interactions~\cite{kokkinogenis2015mobility,rashidi2017exploring}. In the work by Ulloa et al.~\cite{ulloa2016mining}, authors proposed a domain-agnostic framework capable of extracting relevant informations from real-time Twitter data in the field of transportation. 

Kurkcu et al.~\cite{kurkcu2016evaluating} use geo-located tweets to try and discover human mobility and activity patterns. The subject of transport modes was explored by Maghrebi et al.~\cite{maghrebi2016transportation} in the city of Melbourne, Australia. From a dataset of 300,000 geo-located tweets, authors tried to extract tweets related to several modes of transport using a keyword-based search method. 

Additionally, there were also different efforts focused on the tracking of accidents using Twitter social media data. Mai and Hranac~\cite{mai2013twitter} tried to establish a correlation between the California Highway Patrol incident reports and the increased volume of tweets posted at the time they were reported. On the other hand, Rebelo et al.~\cite{rebelo2015twitterjam} implemented a system capable of extract and analyse events related to road traffic, coined  TwitterJam. In that study, authors also used geo-located tweets that were already confirmed as being related to events on the roads and compared their counts with official sources. 

Some authors have also worked towards extracting travel-related content from tweets. For instance, Carvalho et al.~\cite{carvalho2010real} created a travel-related classifier, whose training set had the particularity of being unbalanced, since the percentage of tweets known to be travel-related was very low. Authors reported that using a bootstrapping strategy with linear Support Vector Machines (SVM) in combination with vectorized messages, such as unigram bag-of-words, the performance results were increased. On another work, Kuflik et al.~\cite{kuflik2017automating} proposed a framework to automatically extract and analyse transport-related tweets, in which the filter process was carried out through classification models.

\section{Data}\label{data}
\begin{figure}[b]
\centering
\begin{minipage}{.6\textwidth}
  \centering
  \includegraphics[width=0.9\linewidth]{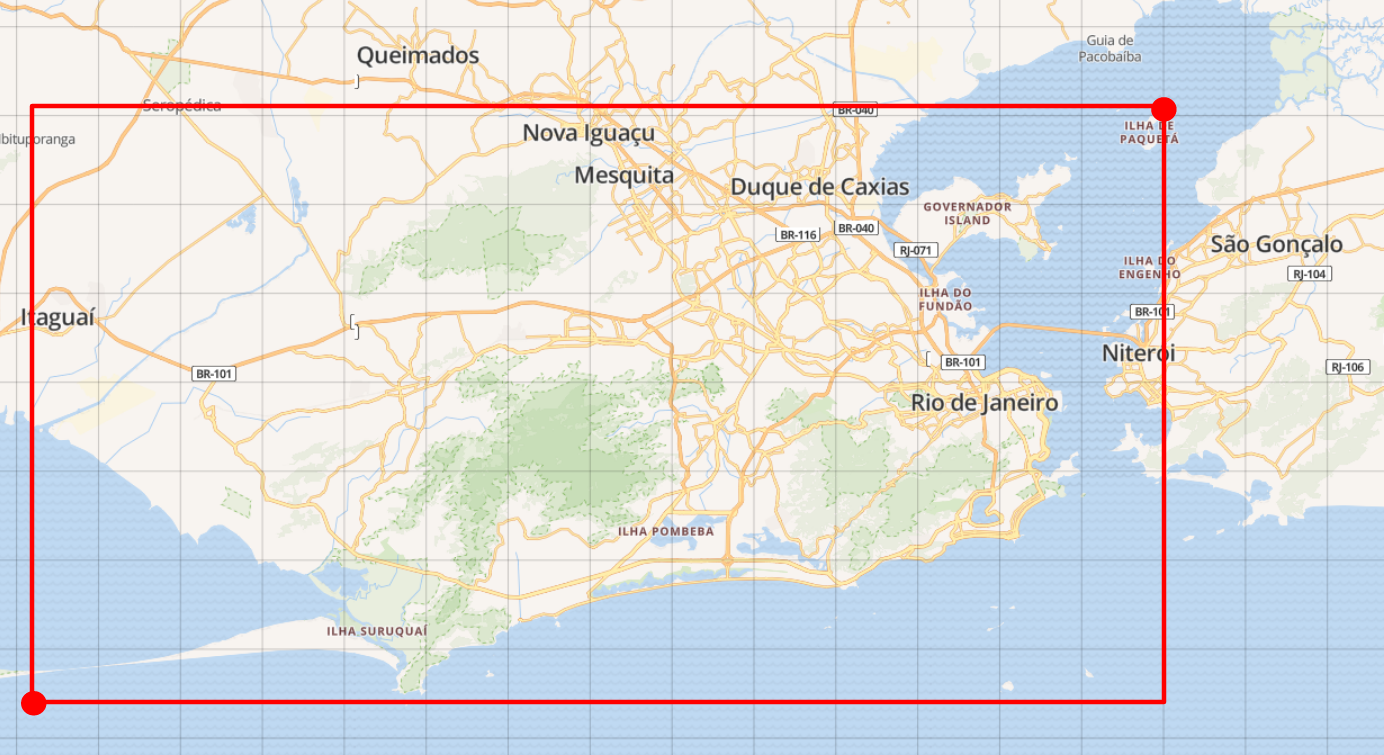}
  \caption{Rio de Janeiro}
  \label{bb_rio}
\end{minipage}%
\begin{minipage}{.4\textwidth}
  \centering
  \includegraphics[width=0.95\linewidth]{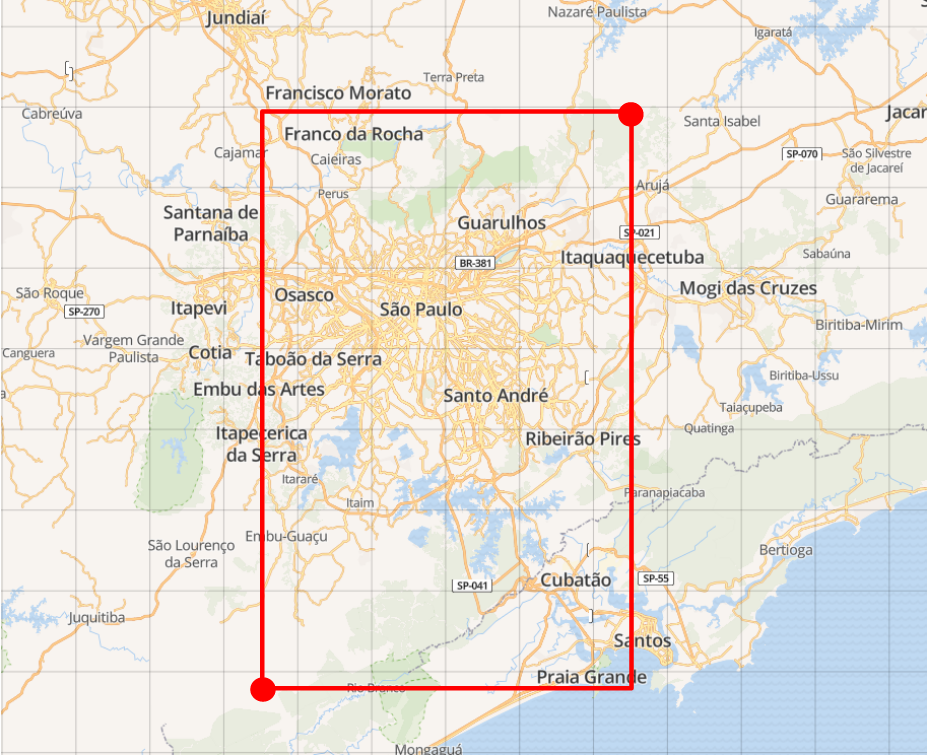}
  \caption{São Paulo}
  \label{bb_sp}
\end{minipage}
\end{figure}

The data used for this study was collected using Twitter's Streaming API through a Python library ( Tweepy\footnote{\url{http://www.tweepy.org/}}). The library was configured with the 'locations' filter activated, whose main purpose is to allow the retrieval of all tweets within a defined bounding-box. Such a bounding-box is a rectangle obtained by two coordinate pairs (latitude and longitude, for the South-West point and the North-East point), as illustrated in Fig. \ref{bb_rio} and in Fig. \ref{bb_sp}. The selected target scenarios in this study are the two largest and most active Brazilian cities on Twitter, namely Rio de Janeiro and São Paulo, both capitals of states with the same names, abbreviated by RJ and SP, respectively.

Messages were collected for a period of one whole month, between days March 12 and April 12, 2017, and the resulting datasets sum up a total of 5.3M and 2.4M tweets for RJ and for SP, respectively.

\subsection{Collecting Geo-located Tweets}
\begin{table}[t]
\small
\centering
\caption{Datasets composition ($10^6$)}
\label{datasets}
\resizebox{\textwidth}{!}{\begin{tabular}{|c|c|c|c|c|c|c|}
\hline
\textbf{City}  & \textbf{All} & \textbf{PT} & \textbf{Non-PT} & \textbf{\begin{tabular}[c]{@{}c@{}}Inside \\ Bounding-Box\end{tabular}} & \textbf{\begin{tabular}[c]{@{}c@{}}Outside\\ Bounding-Box\end{tabular}} & \textbf{\begin{tabular}[c]{@{}c@{}}PT and Inside \\ Bounding-Box\end{tabular}} \\ \hline
Rio de Janeiro & 6,175        & 5,355       & 0,819            & 4,327                                                               & 1,848                                                               & 3,749                                                                      \\ \hline
São Paulo      & 2,934        & 2,444       & 0,490            & 2,016                                          & 0,918                                                                & 1,672                                                                      \\ \hline
\end{tabular}}
\end{table}
As we referred above, the data was collected using a bounding-box-based filter; however, we detected that not all tweets were located inside the defined bounding-boxes. Indeed, the documentation of the Twitter Streaming API mentions that the locations filter has two possible heuristics\footnote{\url{https://dev.twitter.com/streaming/overview/request-parameters#locations}}: (1) if the coordinates field is populated, the values there will be tested against the bounding-box; (2) if the coordinates field is empty but place is populated, the region defined in place is checked for intersections against the locations bounding-box. Any overlapping areas will yield a positive match.

The first heuristic only happens if a user is able/willing to tag a post with his precise geo-location associated with it; otherwise, the user can tag the post associated with a place, selected from a list provided by Twitter, in which case the second heuristic applies. 
Considering the \textit{place} field in a tweet is composed by a bounding-box, if any piece of it overlaps the bounding-box used in the filter process, then a positive match is yielded and the tweet is retrieved. For example, if a tweet has a place such as Brazil and our filter bounding-box is defined for Rio de Janeiro, all tweets from place Brazil will be in our dataset, regardless the fact some tweets are posted elsewhere, such as in the city of Manaus, very far away from Rio de Janeiro.
To solve this matching problem, we used the default Twitter bounding-boxes for RJ and SP to check whether a certain place is or is not inside the desired area. For tweets which field \textit{coordinates} was empty, we calculate the centre of the bounding-box and use that  Only tweets whose field \textit{language} was Portuguese (PT) were considered. The final composition of our dataset is presented in Table \ref{datasets}.

\subsection{Exploratory Data Analysis}
In order to gain better insight into the composition of our datasets, several analysis were performed so as to understand what were the most active hours of the day and days of the week in terms of geo-located tweets, as well as the most mentioned \textit{hashtags} and used words in their contents. This exploratory analysis was performed over the tweets in Portuguese (7.8M tweets).

\begin{figure}[htbp]
\centering
\begin{subfigure}{0.49\textwidth}
\centering
\includegraphics[width=1.0\linewidth]{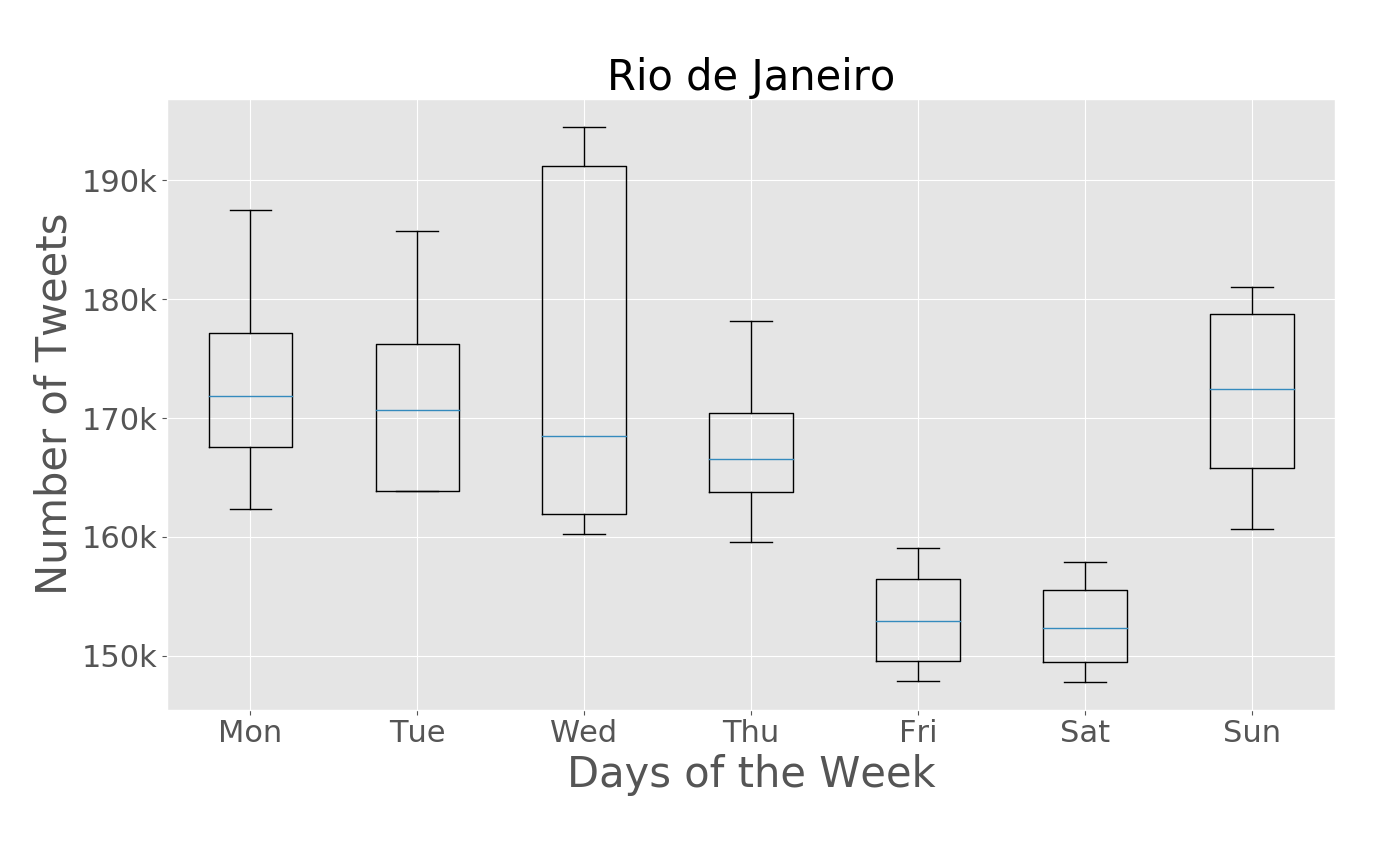}
\label{fig:rio_box_plot}
\end{subfigure}
\begin{subfigure}{0.49\textwidth}
\centering
\includegraphics[width=1.0\linewidth]{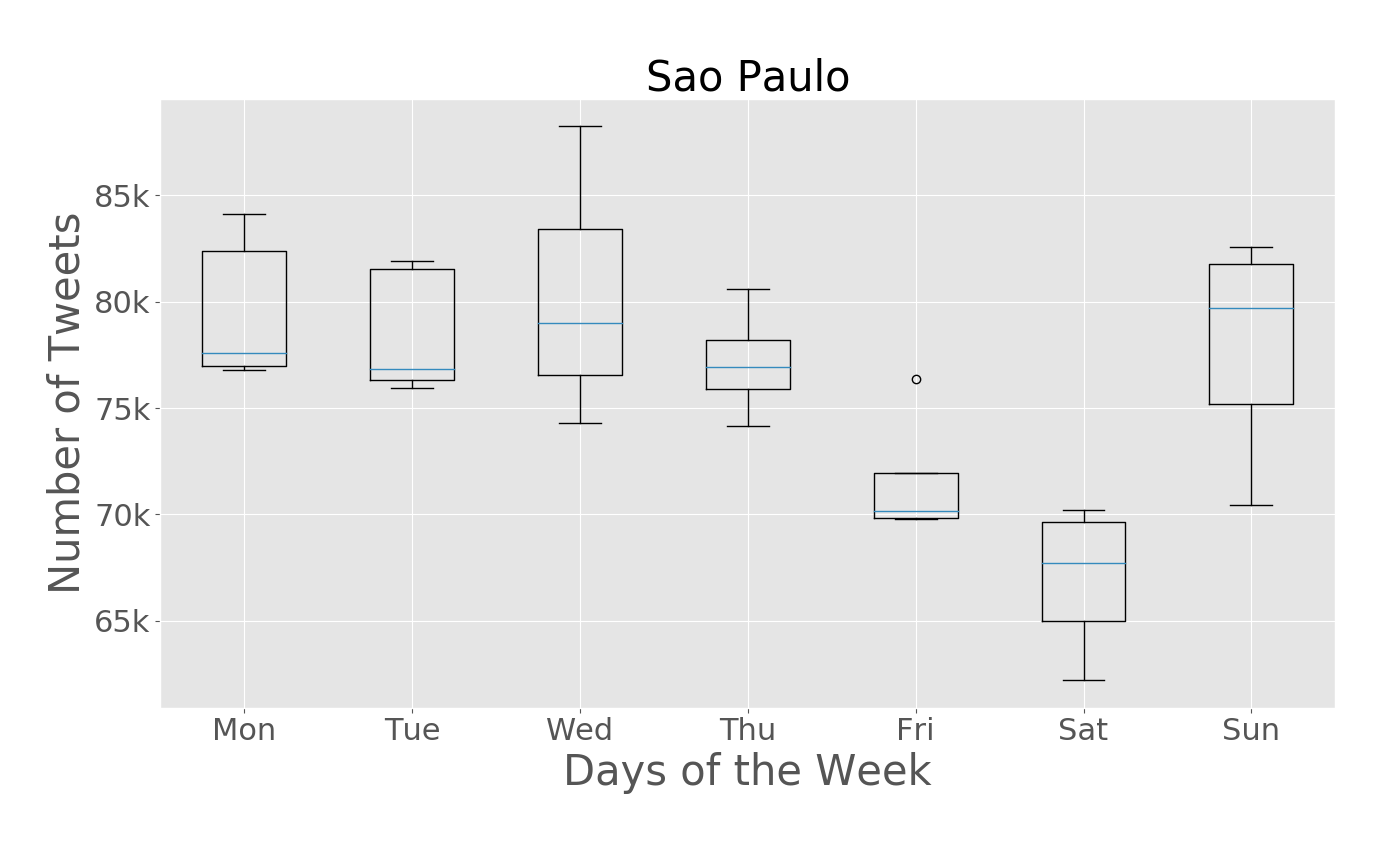}
\label{fig:sp_box_plot}
\end{subfigure}
\vspace{-15pt}
\caption{Frequency of tweets for each day of the week}
\label{box_plot_rio_all}
\end{figure}

Fig. \ref{box_plot_rio_all} illustrates the volume of tweets per day of the week collected during the month of the study. An interesting point to enhance is the difference between the volumes of geo-located tweets of the two cities. According to IBGE (\emph{Instituto Brasileiro de Geografia e Estatística}) statistics, Rio de Janeiro\footnote{\url{http://cod.ibge.gov.br/493}} has 6.5M habitants, half of the population of the city of São Paulo\footnote{\url{http://cod.ibge.gov.br/E4X}} which has 12M citizens. 

On the other hand, our datasets of geo-located tweets comprises 98K distinct users posting geo-located tweets in Rio de Janeiro, while São Paulo only has 78k. In terms of total geo-located tweets, Rio de Janeiro doubles the number of tweets in São Paulo. Such difference in the volume of the datasets can arise several questions about why geo-located Twitter activity in RJ is more intense than SP. 

\begin{figure}[t]
\centering
\begin{subfigure}{0.49\textwidth}
\centering
\includegraphics[width=1.0\linewidth]{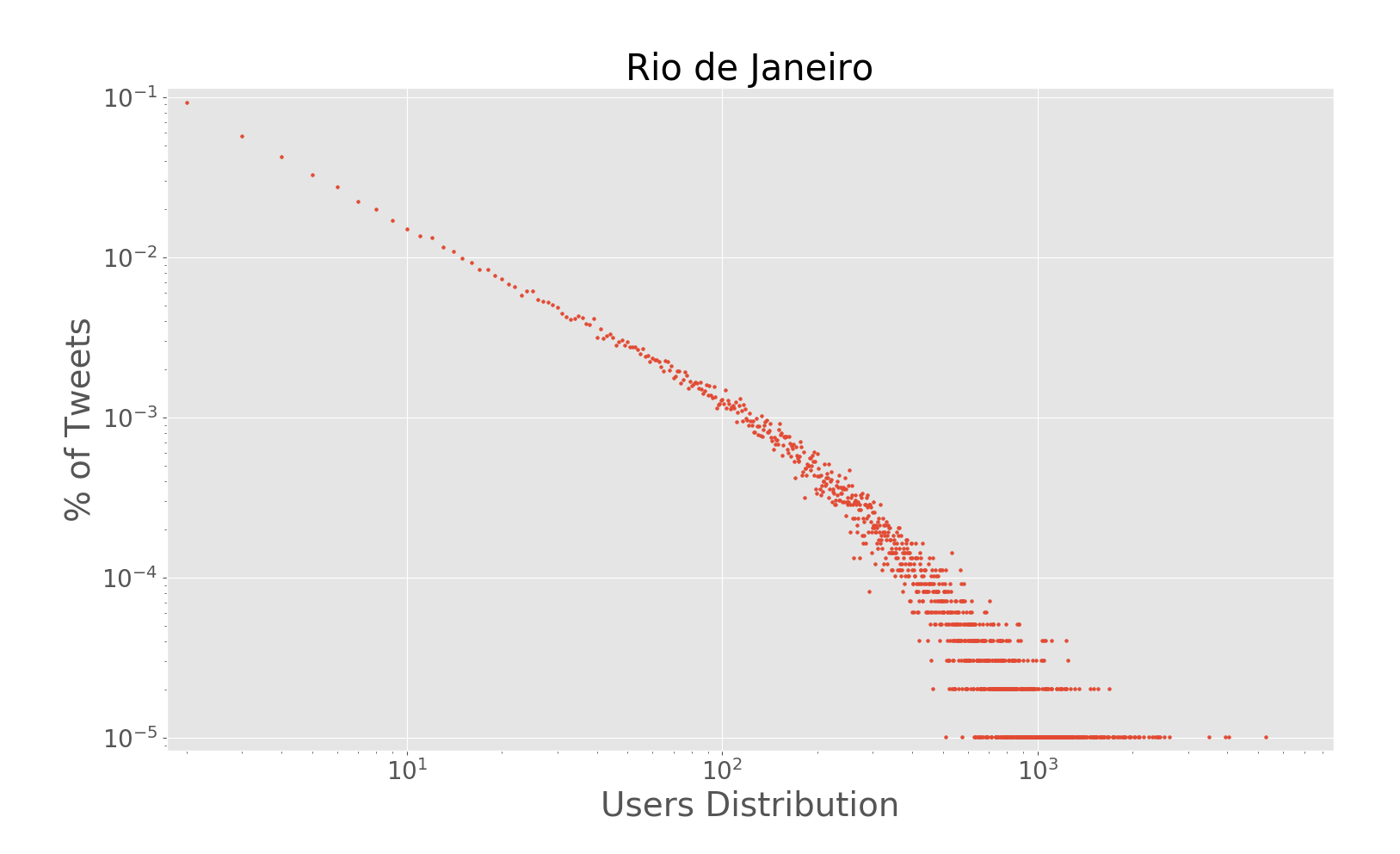}
\label{fig:rio_loglog_plot}
\end{subfigure}
\begin{subfigure}{0.49\textwidth}
\centering
\includegraphics[width=1.0\linewidth]{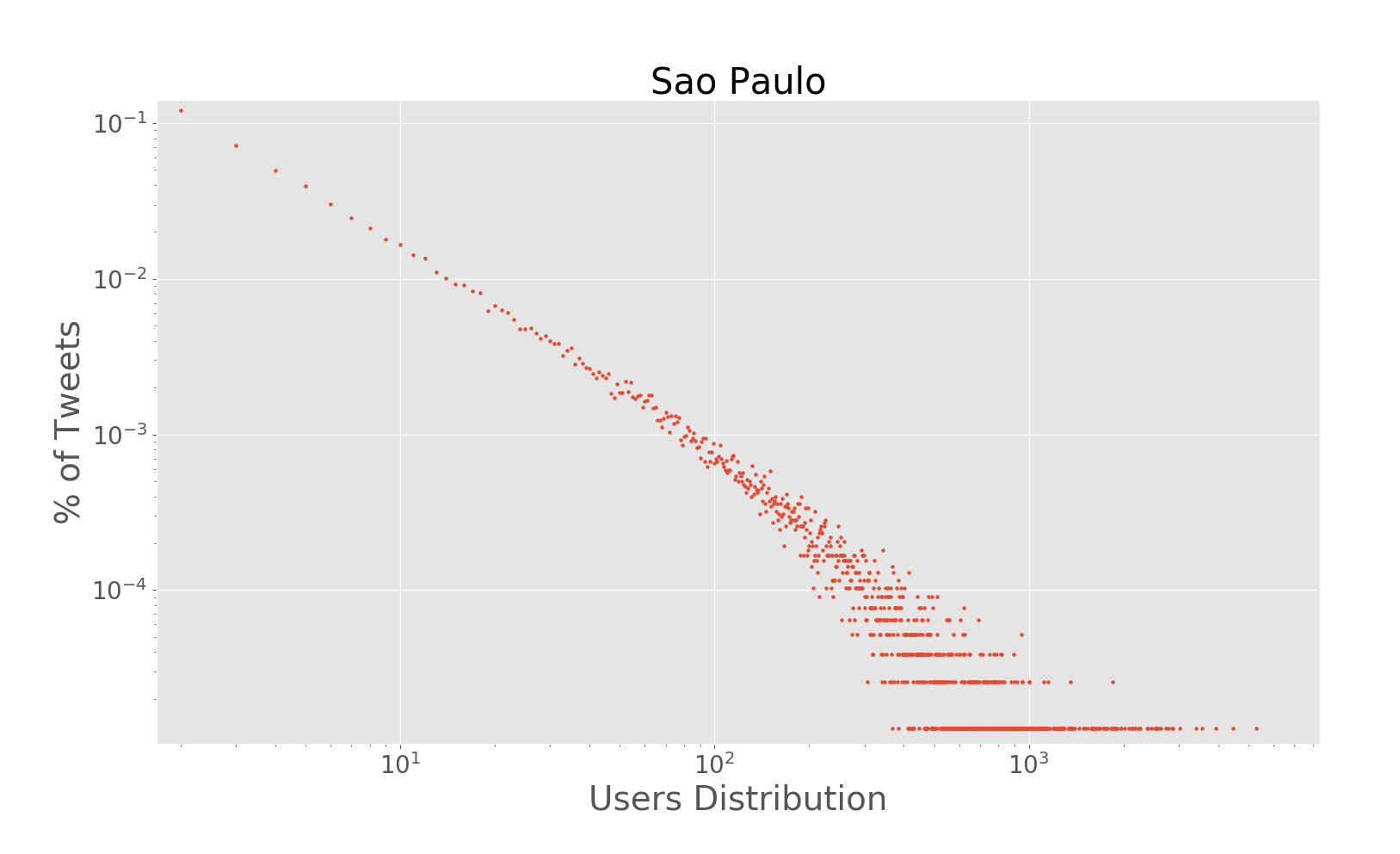}
\label{fig:sp_loglog_plot}
\end{subfigure}
\vspace{-15pt}
\caption{Distribution of users per number of tweets}
\label{loglog_plot_users}
\end{figure}

As we observe in Fig.~\ref{loglog_plot_users}, it is possible to establish correlation between the Power law distribution and the frequency of tweets per user in our datasets. The notable long tail shows us the existence of a large number of occurrences of users whose activity on Twitter is much lower. Such correlation is strongly supported by the number of users that have less than 10 tweets posted in the month of data - 51.7k and 51.5k for RJ and SP, respectively. This values represent more than 60\% of the users in both cities, while 35\% of the remaining users belong to the interval of users that have posted 10-100 tweets. Only a few percentage of users had posted more than 100 tweets.

\section{Classifying Travel-related Tweets}\label{classifier}

The main goal of this work is to support the characterisation of travel-related tweets in Rio de Janeiro and in São Paulo. Considering the volume of the collected data, it was then necessary to automatically identify tweets whose content somehow suggests to be related to the transportation domain. Conventional approaches would require us to specify travel-related keywords to classify such tweets. On the contrary, our approach consisted in training a classifier model to automatically discriminate travel-related tweets from non-related ones. 

One big challenge always present in text analysis is the sparse nature of data, which is especially the case in Twitter messages.
Conventional techniques such as Bag-of-Words tend to produce sparse representations, which become even worse when data is composed by informal and noisy content.

Word embeddings, on the other hand, is a text representation technique that tries to capture syntactic and semantic relations from words. The result is a more cohesive representation where similar words are represented by similar vectors. For instance, \emph{"taxi"/"uber"}, \emph{"bus/busão/ônibus"}, \emph{"go to work"/"go to school"} would yield similar vectors respectively.
We are particularly interested in exploring the characteristics of word embeddings techniques to understand to which extent it is possible to improve the performance of our classifier to capture such travel-related expressions.

\subsection{Data Preparation}
Each tweet of our training and test sets was submitted to a small and basic group of pre-processing operations, as detailed below.

\begin{itemize}
\item \textbf{Lowercasing:} Every message presented in a tweet was converted into lowercase;
\item \textbf{Transforming repeated characters:} Sequences of characters repeated more than three times were transformed, e.g. "loooool" was converted to "loool";
\item \textbf{Cleaning:} Removing URLs and user mentions.
\end{itemize}

\subsection{Features}\label{features}

We established the use of different groups of features to train our classification model, namely bag-of-words, bag-of-embeddings and both combined. Such groups are detailed below.

\begin{itemize}
\item \textbf{Bag-of-words (BoW):} This group of features was obtained using unigrams with standard bag-of-words techniques. We considered the 3,000 most frequent terms across the training set excluding the ones found in more than 60$\%$ of the documents;

\item \textbf{Bag-of-embeddings (BoE):} We created bag-of-embeddings using \textit{paragraph2vec} ~\cite{le2014distributed}. This method is capable of learning distributed representations of words, each word being represented by a distribution of weights across a fixed number of dimensions. Authors have also proved~\cite{mikolov2013linguistic} that this kind of representation is robust when encoding syntactic and semantic similarities in the embedding space. We trained 10 iterations over the whole Portuguese dataset using a context window of value 2 and feature vectors of 100 dimensions. We then took the corresponding embedding matrix to yield the group of features fed into our classification routine. 

\item \textbf{Bag-of-words plus Bag-of-embeddings:} We horizontally combined both the above matrices into a single one and used it as a single group of features.
\end{itemize}

\section{Experimental Setup}\label{exp_setup}
In this section we describe the strategy chosen to create the training and test sets in order to conduct our study. The experimental classification models, their tuning parameters, as well as the evaluation metrics to measure their performance are also discussed.

\subsection{Training and Test Sets}

The construction of the training and test sets followed a traditional approach. We thus tried to select balanced training sets, to which it was necessary to identify tweets that could possibly be travel-related.
We were inspired by a strategy used in the study by Maghrebi~et~al.~\cite{maghrebi2016transportation}, which consists in searching tweets from a collection using specific travel terms and regular expressions. Using the correspondent terms for each mode of transport - (Bike) \textit{bicicleta}, \textit{moto}; (Bus) \textit{onibus}, \textit{ônibus}; (Car) \textit{carro}; (Taxi) \textit{taxi, táxi}; (Train) \textit{metro}, \textit{metrô}, \textit{trem}; (Walk) \textit{caminhar} - combined with the regular expression $space + term + space$, we found about 30,000 tweets. From this subset, we randomly selected a small sample of 3,000 tweets to manually confirm they were indeed related to travel topics. After this manual annotation we selected 2,000 tweets and used them as positive samples in the training dataset.

In order to select negative samples for the training dataset we randomly selected 2,000 tweets and also manually verified their content to assure that they were not travel-related. Finally, our training set was composed by 4,000 tweets, from which 2,000 were travel-related and 2,000 were not. 
We selected 1,000 tweets randomly that were not present in the training set so as to build the test set, and then manually classified them as travel-related or non-travel-related. In the end, 71 tweets were found to be travel-related and whereas 929 were not.

\subsection{Classification}

Support Vector Machines (SVM), Logistic Regression (LR) and Random Forests (RF) were the classifiers used in our experiences. The SVM classifier was tested under three different kernels, namely \textit{rbf}, \textit{sigmoid} and \textit{linear}; the latter proved to obtain the best results. 

The LR classifier was used with the standard parameters, whereas the RF classifier used 100 trees in the forest. The gini criterion and the maximum number of features were limited to those as aforementioned in Section~\ref{features}, in the case of the RF classifier.

\subsection{Evaluation Metrics}

To evaluate the performance of the classifiers in our experiences we used five different metrics. Firstly we compute a group of three per-class metrics, namely precision, recall and the F1-score. Bearing in mind this study considers a binary classification, metrics were associated with the travel-related class only, i.e. the positive class. The ROC (Receiver operating characteristic) curve gives us the TPR (True positive rate) and the FPR (False positive rate) for all possible variations of the discrimination threshold. Through the ROC curve, we compute the area under the curve (AUC) to see what was the probability of the classifier to rank a random travel-related tweet higher than a random non-related one.
\begin{table}[t]
\small
\centering
\caption{Travel-related classifiers results.}
\label{classifiers}
\begin{tabular}{|c|c|c|c|c|}
\hline
\textbf{Classifier}                  & \textbf{Features} & \textbf{Precision} & \textbf{Recall} & \textbf{F1-score} \\ \hline
\multirow{3}{*}{Linear SVM}          & BoW               & 1.0                & 0.6761          & 0.8067            \\
                                     & BoE               & 0.4338             & 0.8309          & 0.5700            \\
                                     & BoW + BoE         & \textbf{1.0}       & \textbf{0.7465} & \textbf{0.8548}   \\ \hline
\multirow{3}{*}{Logistic Regression} & BoW               & 1.0                & 0.6338          & 0.7759            \\
                                     & BoE               & 0.4444             & 0.8451          & 0.5825            \\
                                     & BoW + BoE         & 1.0                & 0.6761          & 0.8067            \\ \hline
\multirow{3}{*}{Random Forest}       & BoW               & 1.0                & 0.6338          & 0.7759            \\
                                     & BoE               & 0.2298             & 0.8028          & 0.3574            \\
                                     & BoW + BoE         & 1.0                & 0.6338          & 0.7759            \\ \hline
\end{tabular}
\end{table}

\section{Results and Analysis}\label{results}

In this section we present the results obtained in the experimental setup. A comparison of different learning algorithms using the features mentioned in Section \ref{features} is provided as so some analysis on tweets classified as travel-related.

\subsection{Results}

Table~\ref{classifiers} presents the results obtained using the different features combination for our test set composed by 1,000 tweets manually annotated. According to the evaluation metrics we conclude that the bag-of-word and bag-of-embeddings combined produced better classification models. The model produced by the Linear SVM performed slightly better than the LR and the RF. Interesting to note is that BoW features have influence on the precision scores obtained from our results, producing more conservative classifiers. Regarding the recall results, we can see that the Logistic Regression using only bag-of-embeddings features was the model with best results; perhaps if the precision is taken into consideration, the same conclusions will not be possible. Analysing the scores provided in Table~\ref{classifiers}, the best model under the F1-score was the Linear SVM, with a score of 0.85.

\begin{figure}[t]
  \centering
  \includegraphics[width=.7\textwidth]{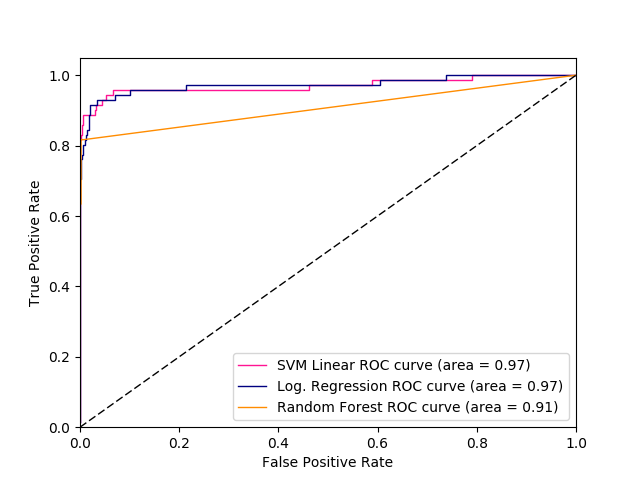}
  \caption{ROC Curve of SVM, LR and RF experiences}
  \label{fig:roc_curve}
\end{figure}

\subsection{Analysis}
\begin{figure}[t]
  \centering
    \includegraphics[width=0.7\textwidth]{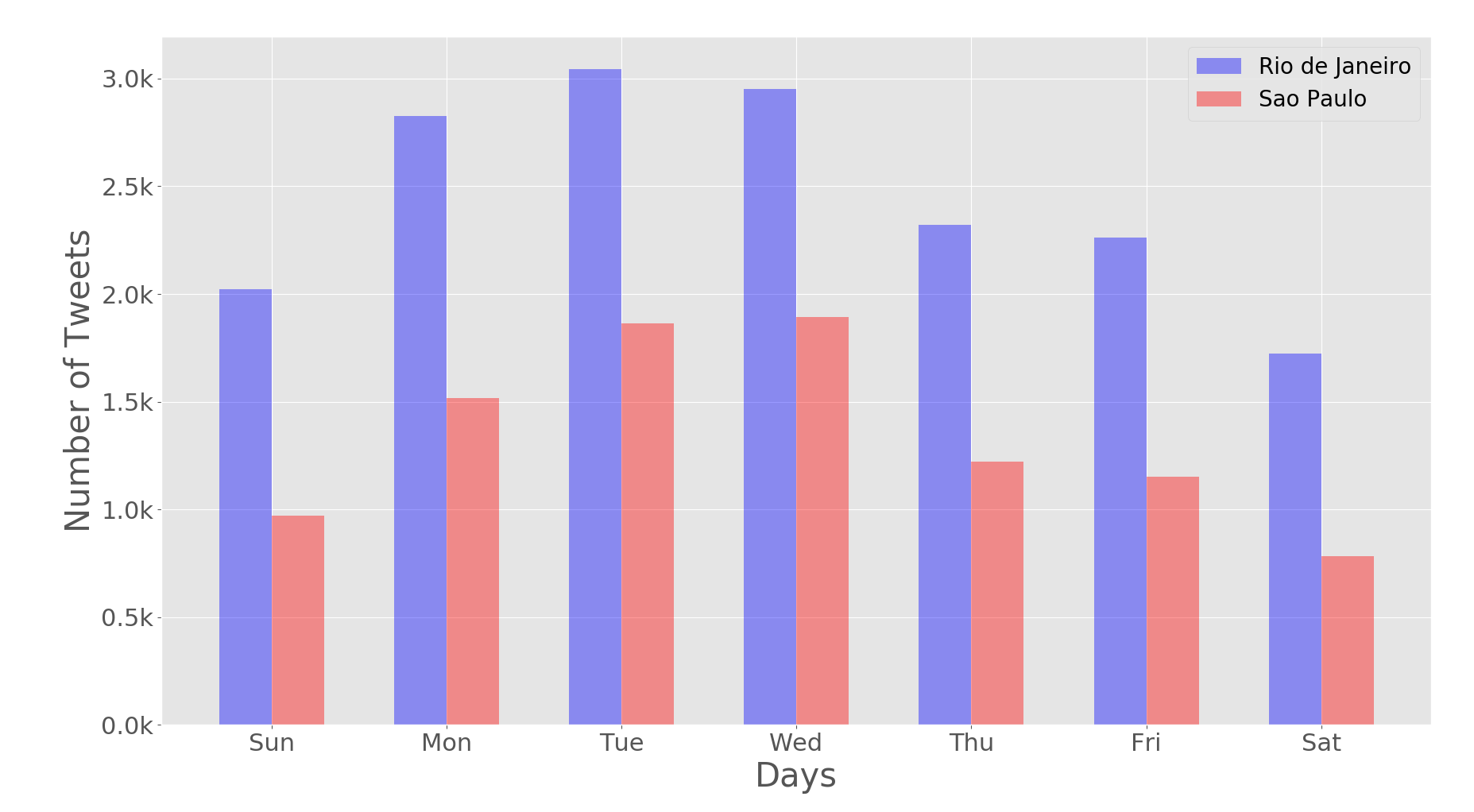}
    \caption{Positive Predicted Tweets per Day of Week}
    \label{predicted}
\end{figure}

The performance of all three classifiers is illustrated using the ROC Curve in Fig. \ref{fig:roc_curve}. The area under the curve of the Receiver Operating Characteristic (AUC) was very similar for both the Logistic Regression and the Linear SVM models. The results obtained from the Random Forest model were not so promising as expected.

\begin{figure}[t]
\centering
\includegraphics[width=0.725\textwidth]{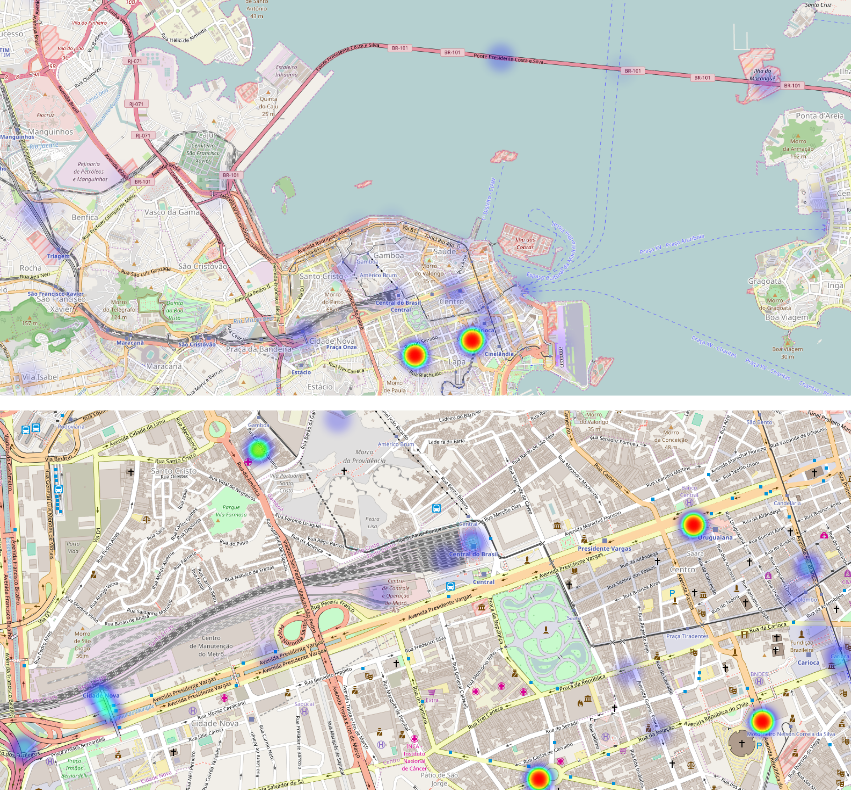}
\caption{Rio de Janeiro Heatmap of the positive tweets}
\label{fig:rio_heatmap}
\end{figure}

After the selection of our classification model, we decided to classify all the Portuguese dataset and draw some statistics from the results. The trained Linear SVM classifier was used to predict whether tweets were travel-related or not, since it was the model presenting the best score under the F1-score metric (as shown in Table~\ref{classifiers}). From a total of 7.8M tweets, our classifier was able identified 37,300 travel-related entries.

Fig.~\ref{predicted} depicts the distribution of travel-related tweets over the days of the week. We can see that the first three business days (Monday, Tuesday and Wednesday) are the ones on which the Twitter activity is higher for both cities.

In order to understand the spatial distribution of travel-related tweets we generated a heatmap for both cities. We calculate the centre of the bounding-box attached to the tweets which field \textit{coordinates} was empty. From the heatmap of RJ, illustrated in Fig.~\ref{fig:rio_heatmap}, it is possible to identify that some agglomerations of tweets are located at Central do Brasil, Cidade Nova and Triagem train stations, as well as at Uruguaiana, Maracanã and Carioca metro stations. The Rio-Niterói bridge, connecting Rio de Janeiro to Niterói, as well as the piers on both sides also presented considerable clouds of tweets classified as travel-related.

The heatmap for the city of SP was also an interesting case to observe. Almost every agglomeration matched some metro or train station. Estação Brás, Tatuapé, Belém, Estação Paulista, Sé, Liberdade were some of the stations highlighted in the heatmap. We could also identify a little agglomeration of travel-related tweets at Congonhas airport, even though no tweets seemed to mention the word \textit{plane} explicitly in the training of our classification model.

\section{Conclusions and Future Work}\label{conclusions_future_work}
The methodology reported across our study is concerned with the problem of the construction of a fine-grained Twitter training set for the travel domain and also the automatic identification of travel-related tweets from a large scale corpus.
We combined different word representations to verify whether our classification model could learn relations between words at both syntactic and semantic levels. After using standard techniques such as bag-of-words and bag-of-embeddings, we have used them combined yielding results that showed that these different groups of features can complement each other.

The future directions for our research will include the application of unsupervised topic modelling algorithms to the non-related subset given by our classifier. 
This extension to our approach will validate the classification performance of our model by analysing different non-travel-related content within the subset, making it possible to identify topics such as tourism, business, night life, among others. We also plan to experiment with more sophisticated preprocessing routines and analysis to improve the quality of our data so as to verify what effect these preparation phases will have on the results obtained.

Another important work to pursue in the future is to correlate the results of this study with official sources of transportation agencies relatively to traffic congestions and other events on the transportation network, including all modes of transports and their integration interfaces. This kind of association will be useful both to validate the proposed approach as well as to improve the inference process and knowledge extraction. The automatic classifier herein presented will then be integrated into data fusion routines to enhance transportation supply and demand prediction processes alongside other sensors and sources of information.

\section*{Acknowledgements}
This work was partially supported by Project "TEC4Growth - Pervasive Intelligence, Enhancers and Proofs of Concept with Industrial Impact/NORTE-01-0145-FEDER-000020".

\bibliographystyle{unsrt}
\bibliography{refs}

\end{document}